\documentclass[12pt]{elsarticle}
\usepackage[T2A]{fontenc}
\usepackage{amsmath}
\usepackage{amssymb}
\usepackage{graphicx}

\renewcommand{\le}{\leqslant}
\renewcommand{\ge}{\geqslant}
\renewcommand{\phi}{\varphi}
\renewcommand{\epsilon}{\varepsilon}

\DeclareMathOperator{\sech}{sech}

\begin{document}
\begin{frontmatter}
\title{Chaotic mixing and fractals in a geophysical jet current}
\author{M.V.~Budyansky, S.V.~Prants}
%
%
\address{Laboratory of Nonlinear Dynamical Systems,
V.I.Il'ichev Pacific Oceanological Institute of the Russian Academy of Sciences,
690041 Vladivostok, Russia}
\begin{abstract}

We model Lagrangian lateral mixing and
transport of passive scalars in meandering oceanic jet currents by 
two-dimensional
advection equations with a kinematic stream function  with a time-dependent
amplitude of a meander imposed. The advection in such a model is known to be chaotic
in a wide range of the meander's characteristics. 
We study chaotic transport in 
a stochastic layer and show that it is anomalous. 
The geometry of mixing 
is examined and shown to be fractal-like. The scattering 
characteristics (trapping time of advected particles and the number 
of their rotations around elliptical points)  
are found to have a hierarchical fractal structure as functions  
of initial particle's positions. A correspondence between the 
evolution of  material lines in the flow and elements of the fractal 
is established. 
\end{abstract}
\begin{keyword}
Chaotic advection\sep meandering jet\sep fractals
\PACS 47.52.+j\sep 47.53.+n \sep 92.10.Ty
\end{keyword}
\end{frontmatter}

\section{Introduction}

Major western boundary currents in the ocean are meandering jets 
separating water masses with different  physical and biogeochemical 
characteristics. The prominent examples are the Gulf Stream in the 
Atlantic Ocean and the Kuroshio in the Pacific Ocean. These and similar 
"heat engines" define the climate in large regions of the planet. Similar 
jets in the stratosphere play important role in transport and distribution 
of chemical substances.
From the hydrodynamic point of view, they may be considered as jet 
flows with running waves of  different   wave lengths  and phase 
velocities imposed. The simplest kinematic model of such a  flow is 
a two-dimensional jet of an ideal fluid with a given velocity profile 
that is perturbed by an amplitude-modulated wave traveling from 
the west to the east.  The problem of transport and mixing of passive 
scalars in meandering jets has been considered by many authors in 
the context of atmospheric and oceanic physics
\cite{SMS89,B89,S92,SWS93,DM93,WK89,NS97,DW96}.

\begin{figure}[!htb]
\begin{center}
\includegraphics[width=\textwidth,clip]{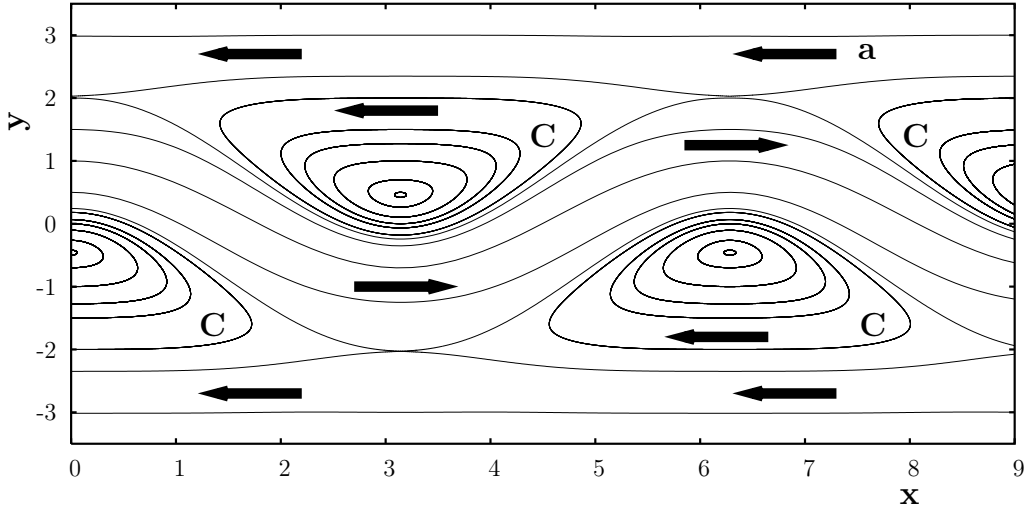}
\caption{Streamlines of the unperturbed system (\ref{11}) in 
the frame moving with the meander's phase velocity $c$.}
\label{fig1}
\end{center}
\end{figure}

The typical phase portrait (Fig.~\ref{fig1}) consists of
two chains of 
circulations with a zigzag-like jet between them and resembles the 
phase portrait of a particle in the field of two running waves. In 
the frame moving with the velocity of one of the waves, the problem 
is topologically equivalent to the motion of a periodically perturbed 
nonlinear physical pendulum that is known to demonstrate chaotic oscillations 
\cite{C79,Z05}. 
Different aspects of chaotic mixing of passive particles in meandering 
jets in the atmosphere and the ocean have been studied in the papers
\cite{SMS89,B89,S92,SWS93,DM93,WK89,NS97,DW96}.

In the paper we focus on topological and statistical aspects of the 
chaotic transport and mixing in a specific kinematic model of an eastward 
meandering jet which has been introduced in Refs. \cite{B89,S92} some years ago. 
We  are motivated by  the desire to get a more deep insight into the 
evolution of material lines in the flow and to establish a connection 
between dynamical, topological and statistical characteristics of the 
flow. The equations of motion of passive particles advected by 
a planar incompressible flow is known to have a Hamiltonian form
\begin{equation}
\begin{aligned}
\dot x&=u(x,y,\,t)=-\frac{\partial \Psi}{\partial y},\\
\dot y&=v(x,y,\,t)=\frac{\partial \Psi}{\partial x},
\end{aligned}
\label{1}
\end{equation}
with the streamfunction $\Psi$ playing the role of a   
Hamiltonian and the coordinates $(x,y)$ of a particle being canonically 
conjugated variables. Thus, nonstationary two-dynamical  advection 
is equivalent to a Hamiltonian system with one and half degrees of 
freedom whose  phase space coincides with its configuration  space. 
This property is very useful in visualizing geometric invariant sets 
in real and numerical experiments.

\section{Model flow}

To be specific we consider a two-dimensional Bickley jet with the 
velocity  profile $u_0\sech^2y$ whose argument is modulated by a zonal 
running wave \cite{S92}. The streamfunction in the fixed frame 
reference is the following:
\begin{equation}
\psi'(x',y',\tau)=-\psi_0\tanh{\left(\frac{y'-a\cos{k(x'-c\tau)}}
{\lambda\sqrt{1+k^2a^2\sin^2{k(x'-c\tau)}} } \right)},
\label{4}
\end{equation}
where $a$, $k$ and $c$  are amplitude, wavenumber and phase velocity 
respectively,  $\lambda$  is  a measure of the jet's width. After introducing 
the following notations: 
\begin{equation}
\begin{aligned}
x&=k(x'-c\tau),
&y&=ky',
&t&=\psi_0 k^2 \tau,\\
x'&=\frac{x}{k}+c\tau,
&y'&=\frac{y}{k},
&\tau&=\frac{t}{\psi_0 k^2}.
\end{aligned}
\label{9}
\end{equation}
and
\begin{equation}
A=ak,\qquad L=\lambda k,\qquad C=\frac{c}{\psi_0 k},
\label{10}
\end{equation}
we get the advection equations  (\ref{1}) in the     frame 
moving with the phase velocity $c$:  
\begin{equation}
\begin{aligned}
\dot x&=\frac{1}{L\sqrt{1+A^2\sin^2 x}\cosh^2\theta}-C,\\
\dot y&=-\frac{A\sin x(1+A^2-Ay\cos x)}{L\left(1+A^2\sin^2 x\right)^{3/2}
\cosh^2\theta},
\end{aligned}\qquad
\theta=\frac{y-A\cos x}{L\sqrt{1+A^2\sin^2 x}}.
\label{11}
\end{equation}

The respective streamfunction
\begin{equation}
\psi(x,y)=-\tanh{\left(\frac{y-A\cos x}{L\sqrt{1+A^2\sin^2 x}}\right)}+Cy
\label{12}
\end{equation}
has three normalized control parameters: $L$, $A$ and $C$ are the jet's  
width, meander's amplitude and its phase velocity. The scaling chosen 
results in translational invariance of the  phase portrait along the  
$x$-axis with the period $2\pi$. 

The detailed analysis of stationary points and bifurcations of Eqs. 
(\ref{11}) has been done in Ref. \cite{ND}. Stationary points may 
exist only under the condition $LC\le 1$. There are four  
stationary points, two of them are always stable and the other ones 
are stable under the condition $AL\operatorname{Arcosh}\sqrt{1/LC}>1$. If the additional 
condition $C<1/L\cosh^2(1/AL)$ is   fulfilled, there are additional 
four unstable saddle points. Resuming one gets:
\begin{enumerate}
\item $C>C_\text{cr1}=1/L$, there are no stationary points.
\item $C_\text{cr1}>C>C_\text{cr2}=1/L\cosh^2(1/AL)$ and 
$C>C_\text{cr3}$, there are two centers and two saddles with  two 
separatrices connecting the saddles. There are two separatrices each of which  
passes through its own saddle. The jet between the separatrices is westward. 
\item $C_\text{cr1}>C>C_\text{cr2}$ and $C<C_\text{cr3}$, the jet is eastward 
with the same stationary points as in the case 2.
\item $C_\text{cr2}>C>C_\text{cr3}$, there are eight stationary points and two separatrices. 
There are two separatrices, each of which connects two
saddle points. The jet is westward.
\item $C_\text{cr2}>C$ and $C<C_\text{cr3}$,  the jet is eastward with the same stationary points 
as in the case 4.
\end{enumerate}

Therefore, from the point of view of existence and stability of stationary
states, there are three possibilities:
1) there are no stationary points; 2) there are four stationary points,
two centers and two saddles; 3) there are eight stationary points, four
centers and four saddles. A bifurcation between the first and second regimes 
consists in arising two pairs ``saddle-center''. A bifurcation between 
the second and third regimes consists in arising two saddles and
a center between them instead of one saddle (a fork-type bifurcation). 

There is one more bifurcation that does not change the number and stability of stationary points but changes the topology of the flow. The values of the streamfunction 
 on the separatrices are equal on modulo but of opposite signs. There is a critical value of the phase velocity $C=C_\text{cr3}$ under which the separatrices  coincide 
and the respective streamfunction is equal to zero. If $C>C=C_\text{cr3}$, a free flow between the separatrices is westward, whereas with $C<C=C_\text{cr3}$ it is eastward.  It is difficult to find $C=C_\text{cr3}$ analytically but it may be shown \cite{ND} that $C_\text{cr3}>C_\text{cr2}$, if 
\begin{equation}
\frac{2(1+A^2)}{AL\sinh(2/AL)}<1.
\label{cond_Ccr3}
\end{equation}
Otherwise $C_\text{cr3}<C_\text{cr2}$. 

The respective phase portraits are typical with Hamiltonian  systems 
with running waves in shear flows \cite{WK89,HH84}. In dependence on the 
values of the phase velocity $C$ one can get different topologies: 
a homoclinic connection, 
a heteroclinic connection and a separatrix reconnection. Being motivated 
by eastward jet currents in the ocean and atmosphere, we deal in this paper 
with the case~3 (see the phase portrait in Fig. 1). 

\section{Chaotic mixing and transport} 

Streamlines with the streamfunction (\ref{12}), that is time-independent in 
the moving frame, are shown in Fig. 1.  The plot demonstrates three 
different regions of the flow: a central eastward jet (J),  north and 
south circulations  (C)  and peripheral westward currents (P). The 
centers of the circulations are at critical lines to be defined by the 
condition  $u(y_c)=c_x$, $v(y_c)=0$, and  they are divided by two 
separatrices connecting 
saddle points. No exchange between the north and south  circulations  
is possible in the unperturbed system. 

Even the simplest periodic  perturbation of the meander's amplitude, 
$A=A_0+ \varepsilon\cos (\omega t + \varphi)$,
results in spitting stable and  unstable manifolds of the saddle points.  
There arise stochastic layers with chaotic mixing and transport of water 
masses fluxes of which depend  on the width of a layer, a number of 
overlapping resonances and other factors. As an example, we plot 
in Fig. 2 the Poincar{\'e} section of the perturbed flow. The control 
parameters throughout the paper are chosen as follows:  $A=0.785$, 
$C=0.1168$, $L=0.628$, $\varepsilon=0.0785$, $\omega= 0.2536$ and 
$\varphi=\pi/2$. 
\begin{figure}[!htb]
\begin{center}
\includegraphics[width=\textwidth,clip]{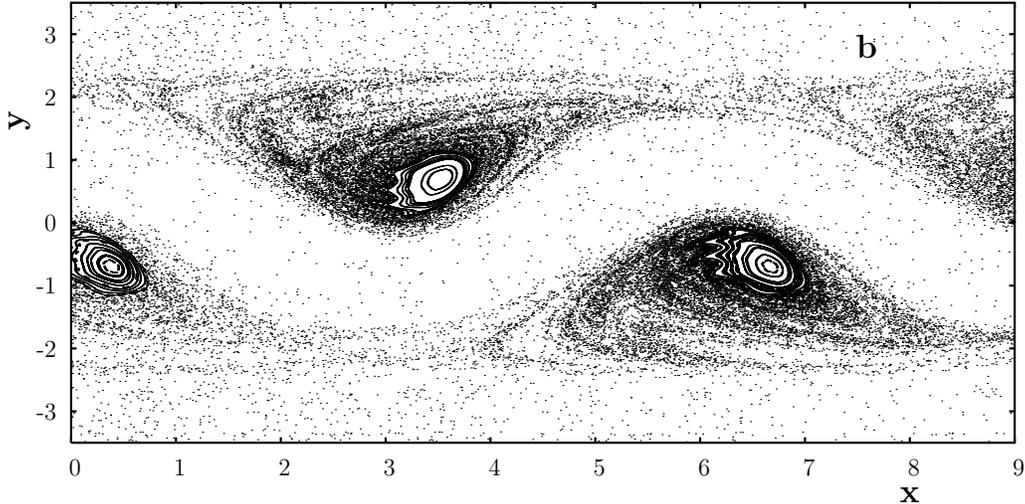}
\caption{General view of the Poincar\'e section of the system with 
a periodically modulated meander's amplitude.}
\label{fig2}
\end{center}
\end{figure}
\begin{figure}[!htb]
\begin{center}
\includegraphics[width=\textwidth,clip]{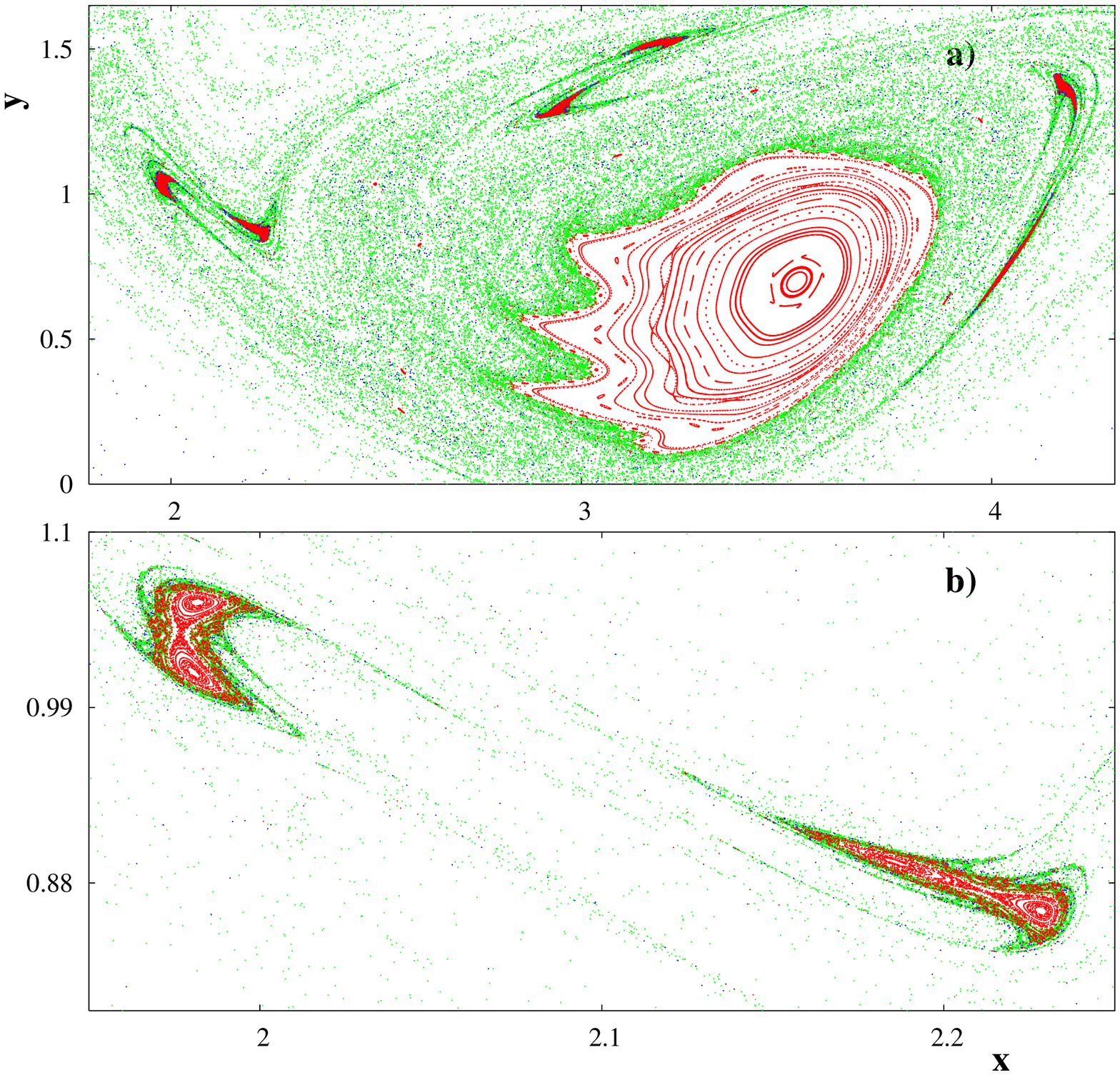}
\caption{(a) Poincar\'e sections in the northern part 
of the first eastern frame, (b) stickiness to the island's border.}
\label{fig3}
\end{center}
\end{figure}

Due to the zonal and meridian symmetries it is enough to consider mixing 
in the northern part of the first eastern frame only $0\le x\le 2\pi$. In Fig. 3a we 
plot the respective Poincar\'e section. The vortex core, that survives 
under the perturbation, is immersed into a stochastic sea where one 
can see 6 small islands belonging to the same resonance. Particles, 
belonging to these islands, rotate around the elliptic point of the vortex core. 
Zoom of the section nearby two of the islands is shown in Fig. 3b.  
The feature we want to pay attention to is a stickiness to the 
boundaries of the vortex core and of the islands that is visualized by increasing 
the density of points near the respective boundaries. It is a typical 
phenomenon with Hamiltonian  systems  \cite {Z05} which influences 
essentially the transport of passive particles. 
\begin{figure}[!htb] 
\centerline{\includegraphics[width=\textwidth,clip]{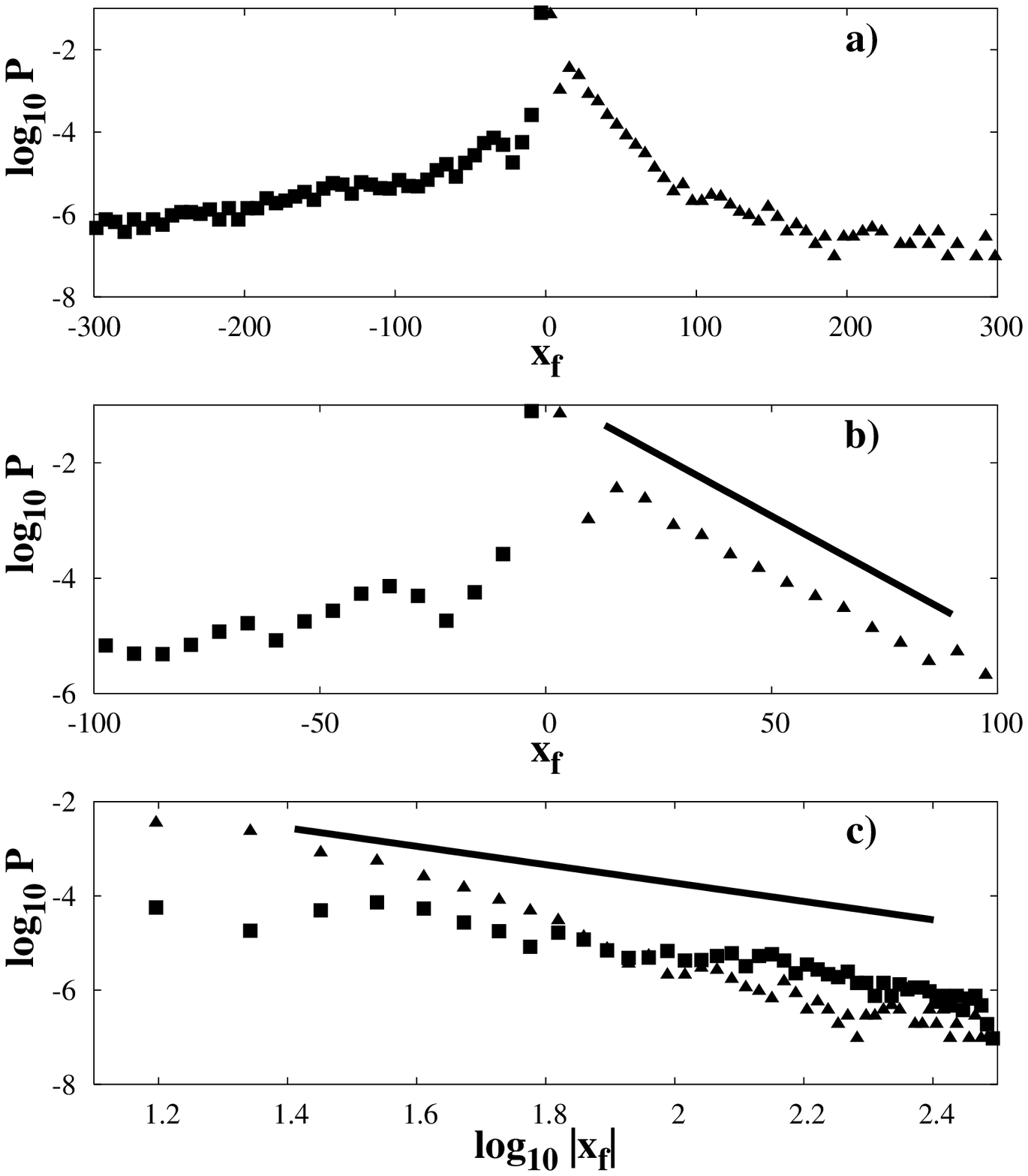}} 
\caption{ 
(a) Probability distribution functions (PDFs) of lengths
of eastward (triangles) and westward (squares) flights
obtained with $10$ chaotic trajectories for $5\cdot 10^6$ steps.
The parameters are: $\varepsilon=0.0785$ and $\omega=0.2536$.
(b) For short eastward flights, $12\le x_f\le 90$, the PDF decays
exponentially. (c) The tails of both
the PDFs decay as a power law with the slope
$\nu=1.94 \pm 0.09$ obtained with westward (squares) long flights $|x_f|>25$.
} 
\label{fig4} 
\end{figure} 
Without perturbation, 
the transport properties are very simple: particles either rotate 
in circulations C or move eastward in the jet J  and westward in 
the peripheral currents P.
Under a perturbation, the motion in the stochastic layers become extremely 
sensitive to small variations in initial conditions, and one is forced 
to use an statistical approach to describe transport. 
A commonly used statistical measure of transport is the variance $\sigma^2(t)
=<x^2>$,  
where the averaging is supposed to be done over an ensemble of particles. 
Long time behavior of $\sigma^2(t)$ and of the probability density function of 
particle's displacement  $P(x,t)$ may be anomalous with typical chaotic 
Hamiltonian  systems  \cite{SZK93}. When  $\sigma^2(t) \sim  t$ and 
$P(x,t)$  is a Gaussian 
distribution, the advection is known to be normal with a well-defined 
diffusion coefficient $D=\lim_{t\to\infty} \sigma^2/2t$. 
When $\sigma^2(t) \sim  t^\gamma$, with $\gamma\ne 1$,  $D$ is either 
zero or infinite \cite{SZK93},
and one gets either subdiffusion ($\gamma<1$) or super-diffusion ($\gamma>1$). 
If the trajectories are dominated by sticking regions nearby boundaries 
of islands, where particles spend a long time, subdiffusion results. 
Superdiffusion occurs when particles in the jet travel long distances 
between sticking events. The respective length and time PDFs are expected to be non Gaussian. 

We will
call ``a flight'' any event between two successive changes of signs
of the particle's zonal velocity. In this terminology
a sticking consists of a number of flights with approximately equal
flight times. 
The Poincar{\'e} section of the flow with the perturbation
strength $\varepsilon=0.0785$ and frequency $\omega=0.2536$ is
shown in Fig.~\ref{fig3}a. The flight PDFs are computed with 
$10$ particles (initially placed in the first east frame inside
a stochastic layer) up to the time $t=5\cdot 10^6$. The PDF
of the lengths of flights is shown in Fig.~\ref{fig4}a for both the
directions. The asymmetry between the eastward (``positive'')
and westward (``negative'') flights is evident. Both the PDFs
can be roughly split into three distinctive regions. The very short
flights with small values of $|x_f|$ ($<2\pi$) are supposed
to be dominated by sticking to the boundaries of the vortex core and
oscillatory islands. The PDF for eastward flights with the lengths in the range
$12\le x_f\le 90$ decays exponentially (Fig.~\ref{fig4}b).
The tails of both the PDFs are close to a power-law decay
$P(x_f)\sim |x_f|^{-\nu}$. In Fig.~\ref{fig4}c we estimate
the exponent for long westward flights, $|x_f|>25$, with corresponding
error by least-square fitting of the straight line 
to the log-log plot of the data to be
$\nu=1.94 \pm 0.09$. 
\begin{figure}[!htb]
\begin{center}
\includegraphics[width=\textwidth,clip]{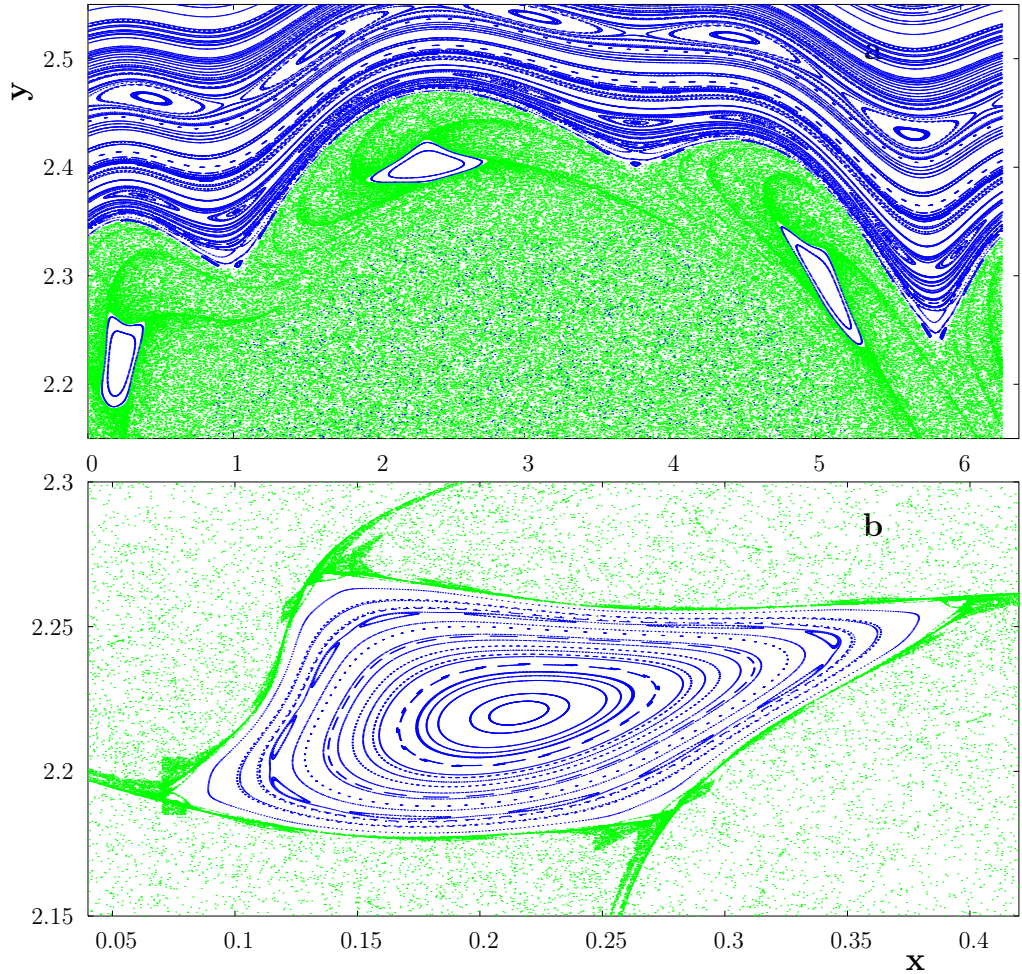}
\caption{(a) The northern border between the circulation (C) 
and the peripheral current (P), (b) stickiness to the border of a ballistic 
island.}
\label{fig5}
\end{center}
\end{figure}
\begin{figure}[!htb]
\begin{center}
\includegraphics[width=0.8\textwidth,clip]{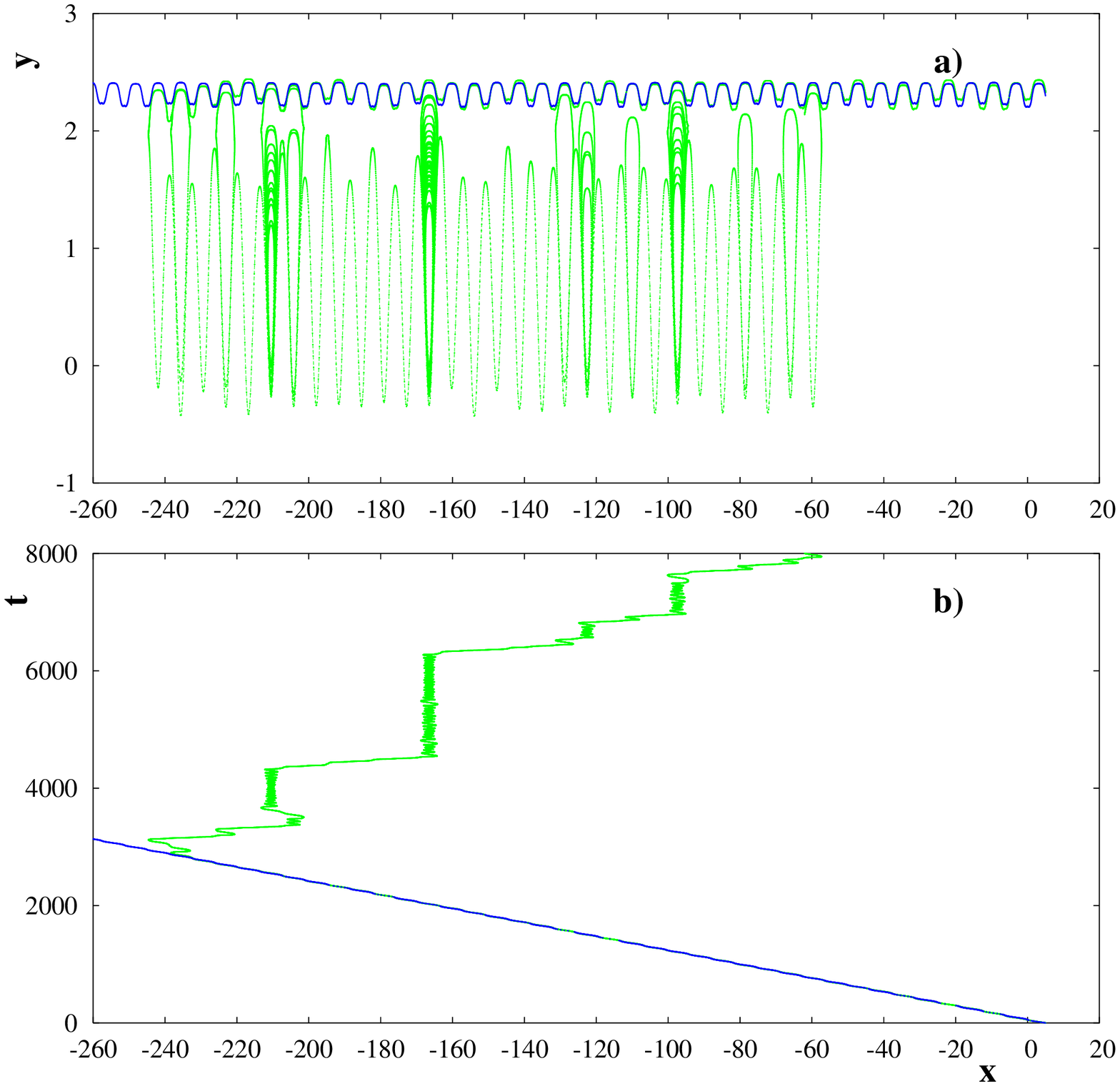}
\caption{Examples of ballistic trajectories: 
the blue and regular trajectory, which is the upper one in (a) 
and the lower one in (b), is inside a ballistic island, the green and 
weakly chaotic trajectory, which is the lower one in (a) 
and the upper one in (b),  is just outside the island. (a) $x-y$ plane, 
(b) dependence of the zonal position on time. Stickiness and flight events 
are evident with the green chaotic trajectory.}
\label{fig6}
\end{center}
\end{figure}

Besides resonant islands with particles moving around the elliptic 
point in the same frame (Fig. 3), we have found so-called ballistic 
islands which were situated both in the  chaotic sea and in the 
peripheral jets (Fig. 5a). Ballistic modes \cite{Z05,VRKZ99,IGF98} 
correspond to the stable periodic motion of particles
from one frame to another. Particles, belonging to a ballistic island, 
move with a constant zonal velocity from one frame to another. When mapping 
their positions at the moments $t=2k\pi/\omega$     $(k=1,2, \dots)$ 
onto the first 
frame, we see a chain of islands which are visible in Fig. 5a. In 
Fig. 5b a zoom of the ballistic island is shown. Stickiness to the 
island's border is evident. 
In Fig. 6 we demonstrate two ballistic trajectories corresponding 
to the ballistic island shown in Fig. 5b. If a particle at $t=0$ is 
placed inside the island, it travels to the west in a regular way 
(the upper blue trajectory in Fig. 6a). Its zonal position $x$ grows 
linearly with time (the lower blue trajectory in Fig. 6b). The lower green 
trajectory in Fig. 6a corresponds to a particle  placed initially 
nearby the border of the same ballistic island from outside. 
Intermittent flight and sticking events are evident in Fig. 6b 
(the upper green curve in that figure). 

\section{Fractal geometry of mixing}
\begin{figure}[!htb]
\begin{center}
\includegraphics[width=0.8\textwidth,clip]{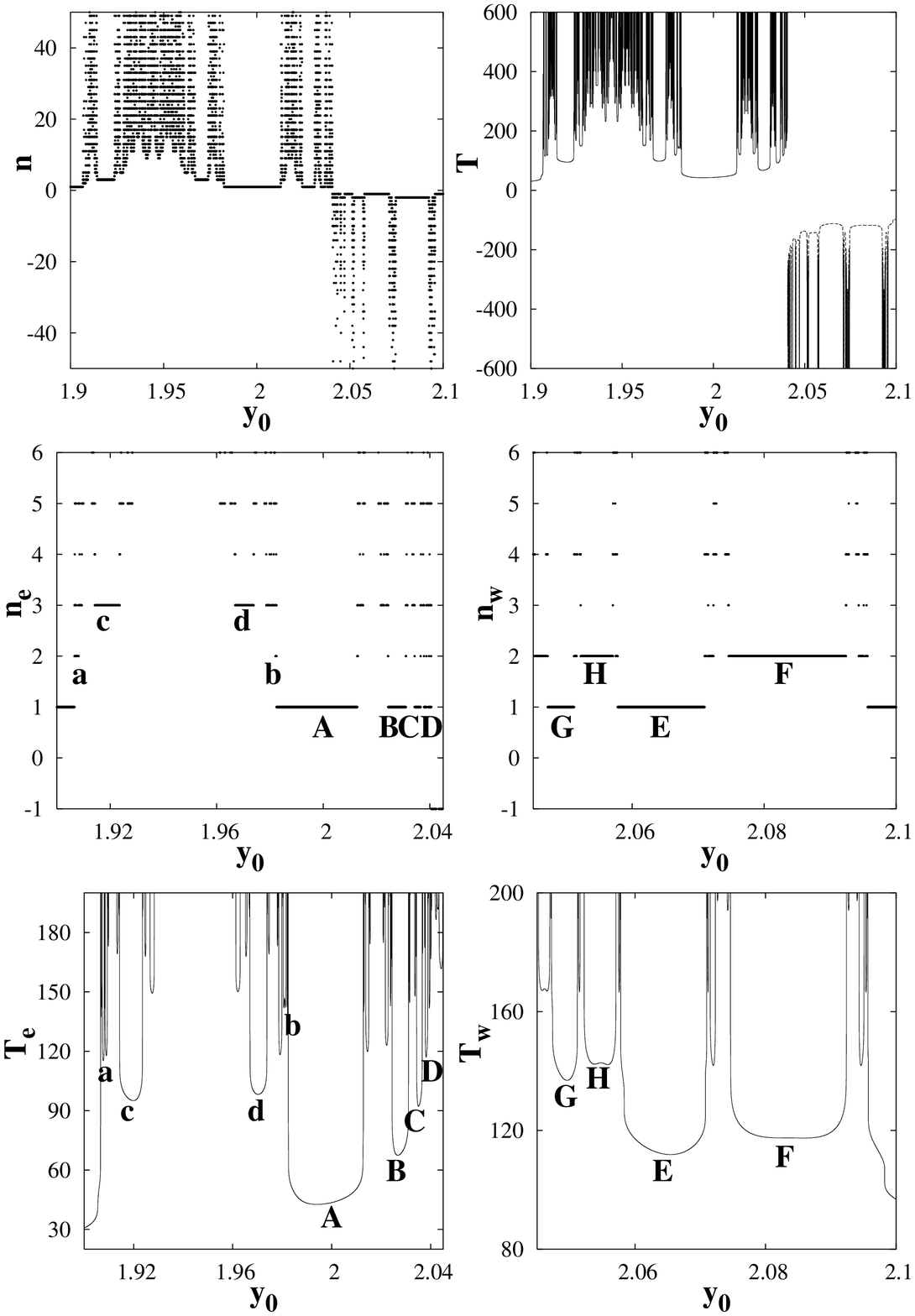}
\caption{Fractal set of initial positions $y_0$ of particles that reach 
the lines $x=0,\pm2\pi$ after $n/2$ turns around the elliptic points. 
$T$ is a time particles need to reach the lines $x=0,\pm2\pi$. 
Indices $e$ and $w$ mean particles moving in the eastward and westward 
directions, respectively.} 
\label{fig7}
\end{center}
\end{figure}
\begin{figure}[!htb]
\begin{center}
\includegraphics[width=\textwidth,clip]{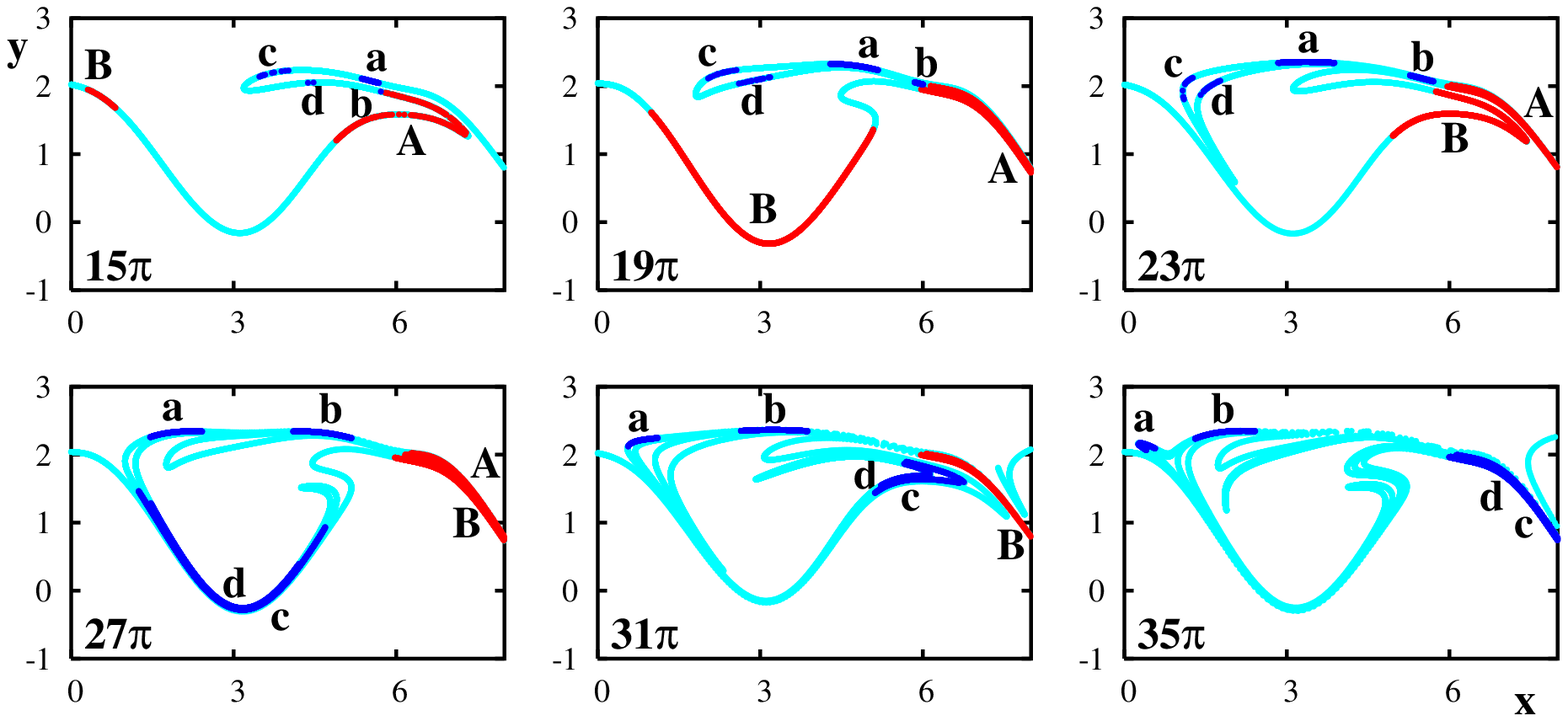}
\caption{Fragments of the evolution of a material line in the first eastern 
frame. The fragments of the fractal in Fig. 7 with $n_e=1,2,3$ are marked by 
the respective letters.} 
\label{fig8}
\end{center}
\end{figure}
Poincar{\'e} sections provide good impression about the structure of 
the phase space but not about geometry of mixing. In this section we consider 
the evolution of a material line consisting of a large number of particles 
distributed initially on a straight line that transverses the stochastic 
layer at $x=0$. A typical stochastic layer consists of an infinite number 
of unstable periodic and chaotic orbits  with islands of 
regular motion to be imbeded. All the unstable invariant sets are 
known to possess  stable and unstable manifolds. When time progresses 
particle's trajectories nearby a stable manifold of an invariant set 
tend to approach the set whereas the trajectories close to an unstable 
manifold go away from the set. Because of such a very complicated 
heteroclinic structure, we expect a diversity of particle's trajectories. 
Some of them are trapped forever in the first eastern frame $0\le x\le 2\pi$
rotating around the elliptic point along heteroclinic orbits. Other ones quit the frame through the 
lines $x=0$ or $x=2\pi$,  and then either are  trapped  there or move to the 
neighbor frames (including the first one), and so on to infinity. 

To get a more deep insight into the geometry of chaotic mixing we follow 
the methodology of our works \cite {PD04,JETP04} and compute the time $T$, 
particles spend in the neighbor circulation zones $-2\pi \le x\le 2\pi$ 
before reaching 
the critical lines $x=0, x=\pm2\pi$, and the number of times $n/2$ they 
wind crossing the  lines $x=\pm \pi$. In the upper panel in Fig. 7 
the functions $n(y_0)$ and $T(y_0)$ are shown. The upper parts of each function  
(with $n>0$ and $T>0$) represent the results for the particles with initial 
positive zonal velocities which they have simply due to their locations 
on the material line at $x=0$. These particles enter the eastern frame 
and may change the direction of their motion many times before leaving 
the frame through the lines $x=0$ or $x=2\pi$. The time moments of those events 
we fix for all the  particles with $1.9 \le y_0 \le 2.045$.
The lower "negative" parts of the functions $n(y_0)$ and $T(y_0)$  represent 
the results for the  particles with initial negative zonal velocities 
($y_0\ge 2.045$) which move initially to the first western frame 
($-2\pi \le x\le 0$). In fact, $T_e(y_0)$ and $T_w(y_0)$ 
 are the time moments when a particle with the initial position 
$y_0$ quits the eastern or western frames, respectively. Both the functions 
consist of a number of smooth U-like segments intermittent with poorly 
resolved ones. Border points of each U-like segments separate particles 
belonging to stable and unstable manifolds of the heteroclinic structure. 
The corresponding initial $y$-positions is a set (of zero measure) of 
particles to be trapped forever in the respective frame. A fractal-like 
structure of chaotic advection in both the 
frames is shown in the upper panel in Fig. 7, and its fragments for the 
first levels are shown in the middle panel for the eastern and the western 
fragments separately. Particles with even values of $n$  
quit one the frames through the border $x=0$, those with odd $n$ -- through 
the border $x=2\pi$ for the eastern frame and $x=-2\pi$  for the western one.

Let us consider in detail the fractal-like structure in the eastern frame 
keeping in mind that the results are similar with any other frame. The 
$n_e(y_0)$-dependence is a complicated hierarchy of sequences of segments of 
the material line. Following to the authors of the paper \cite{Chaos}, 
we call as an epistrophe a sequence of segments of the $(n+1)$-level, 
converging to the ends of a segment of a sequence of the $n$-th level, 
whose length decrease in accordance with a law. At $n_e=1$ we see in 
Fig. 7 an epistrophe with segment's  length A, B, C, D and so on decreasing 
as $l_m=l_0q^m$ with $q \approx 0.46$. 
Letters  a and b in Fig. 7 denote the first segments of the 
epistrophes at the level $n_e=2$, whereas d and c ---the first segments of 
the epistrophes at the level $n_e=3$. The respective laws for all those 
epistrophes are not exponential. 

In Fig. 8 we demonstrate  fragments of  the evolution of the material 
line in the first eastern frame at the moments indicated in the figure. 
Letters 
on the line mark the corresponding segments of the $n_e(y_0)$ and $T_e(y_0)$  
functions in Fig. 7. As an example, let us explain formation of the 
epistrophe ABCD at the level $n_e=1$. With the period of perturbation 
$T_0=2\pi/\omega \simeq 8\pi$, a portion from the north end of the material 
line leaves 
the frame through its eastern border. Look at the segments A and B at 
$t=15\pi$ and $t=23\pi$. They quit the first frame as a fold through 
the period $T_0\simeq 8\pi$. The other segments -- C, and D   (not shown 
in Fig. 8) do 
the same job. The epistrophe's segments at the odd levels ($n=2k-1>1$) quit 
the frame with the period of perturbation $T_0$ one by one  being folded 
(c and d segments). The folds of the segments of the $(2k-1)$-level 
are exterior with respects to the folds of the segments of the $(2k+1)$-level. 
The following empirical law is valid: $T_{2k-1}-T_{2k+1} \simeq 2T_0$, 
where $T_{2k-1}$ is a time 
when the first segments of the epistrophes at the level  $(2k-1)$  
(A with $n_e=1$) reach the line $x=2\pi$, and $T_{2k+1}$ the respective 
time for the first segments 
of the epistrophes at the level  $2k+1$ (c and d segments with $n_e=3$). 

Segments of the epistrophes of the even levels ($n=2k$) leave the  frame 
with the period $T_0$ as well but through the border
$x=0$ moving to the west. We show the evolution of some of them at the 
moments $t=31\pi$ and $t=35\pi$  in Fig. 8. Thus, the material  line 
evolves by stretching and 
folding, and folds quit the frame in both directions with the 
period of perturbation.    

\section{Conclusion}

We have treated the problem of mixing and transport of passive particles in a kinematic model of 
a meandering oceanic jet current from the point of view of dynamical system's theory. 
A careful simulation of the Hamiltonian equations of advection has shown a complicated character of 
mixing under a time-dependent perturbation of the meander's parameters. Both the oscillatory and ballistic 
resonant islands  and sticking of trajectories to their boundaries have been found.  
The transport has benn shown to be anomalous. The geometry  of mixing has been shown to be fractal-like. The trapping time of advected particles and the number of their rotations around elliptical points   
have been found to have a hierarchical fractal structure as functions  
of initial particle's positions. A correspondence between the 
evolution of  material lines in the flow and elements of the fractal 
has been  established. 

\section{Acknowledgments}
This work was supported by the Russian Foundation for Basic Research 
(Project no.06-05-96032), 
by the Program ``Mathematical Methods
in  Nonlinear Dynamics'' of the Russian Academy of Sciences  
and by the Program for
Basic Research of the Far Eastern Division of the Russian Academy of
Sciences.

\end{document}